\documentclass[number, sort&compress]{elsarticle}
\usepackage{amsmath}
\usepackage{amssymb}
\usepackage{amsbsy}
\usepackage{amsgen}
\usepackage{graphicx}
\usepackage{natbib} 

\usepackage{color}	
\usepackage{ulem}   

\newdefinition{propo}{Proposition}
\newdefinition{corol}{Corollary}

\newdefinition{rmk}{Remark}
\newproof{pf}{Proof}

\everymath{\displaystyle}

\begin{document}
\begin{frontmatter} 

\title{On the properties of input-to-output transformations in networks of perceptrons}

\author[ggc,bro,ste]{A.V.~Olypher~\corref{cor1}}
\ead{aolifer@ggc.edu}
\author[uag]{J.~Vaillant}
\ead{Jean.Vaillant@univ-ag.fr}

\cortext[cor1]{Corresponding author}

\address[ggc]{School of Science and Technology, Georgia Gwinnett College, Lawrenceville, GA, USA}
\address[bro]{Department of Physiology and Pharmacology, SUNY, Brooklyn, NY}
\address[ste]{The Stentor Institute, Prague, Czech Republic}
\address[uag]{Department of Mathematics and Informatics, University of the French Antilles and Guyana, Pointe-\` {a}-Pitre, Guadeloupe, France}

\begin{abstract}
Information processing in certain neuronal networks in the brain can be considered as a map of binary vectors, where ones (spikes) and zeros (no spikes) of input neurons are transformed into spikes and no spikes of output neurons. A simple but fundamental characteristic of such a map is how it transforms distances between input vectors.  In particular what is the mean distance between output vectors given certain distance between input vectors?  Using combinatorial approach we found an exact solution to this problem for networks of perceptrons with binary weights.  he resulting formulas allow for precise analysis how network connectivity and neuronal excitability affect the transformation of distances between the vectors of neuronal spiking. As an application, we considered a simple network model of information processing in the hippocampus, a brain area critically implicated in learning and memory, and found a combination of parameters for which the output neurons discriminated similar and distinct inputs most effectively. A decrease of threshold values of the output neurons, which in biological networks may be associated with decreased inhibition, impaired optimality of discrimination.	
\end{abstract}

\begin{keyword}
neuronal networks \sep perceptrons \sep Hamming distance \sep hippocampus
\end{keyword}
\end{frontmatter} 


\section{Introduction}
In many brain areas neuronal spiking does not correlate directly with external stimuli or motor activity of the animal.  Information processing in such areas is poorly understood. Neurons apparently transform abstract inputs to outputs. Revealing the character of those transformation is challenging.  For example, in the hippocampus -- a brain area critically implicated in learning and memory \cite{And06} -- principal neurons are connected with tens of thousands input neurons \cite{Meg01}. The most advanced experimental techniques allow for selective activation of no more than hundred connections ("synapses")  \cite{Kra09}.  In computer simulations of neuronal models arbitrary spatio-temporal input patterns can be considered.  However constraints on computational resources limit sampling of input patterns, and neuronal models, especially in neuronal networks, are most often simplified to decrease the number of possible combinations of inputs \cite{Ney11, Cut09}. 

\begin{figure} [t]
\centering
\includegraphics  [scale=0.4]{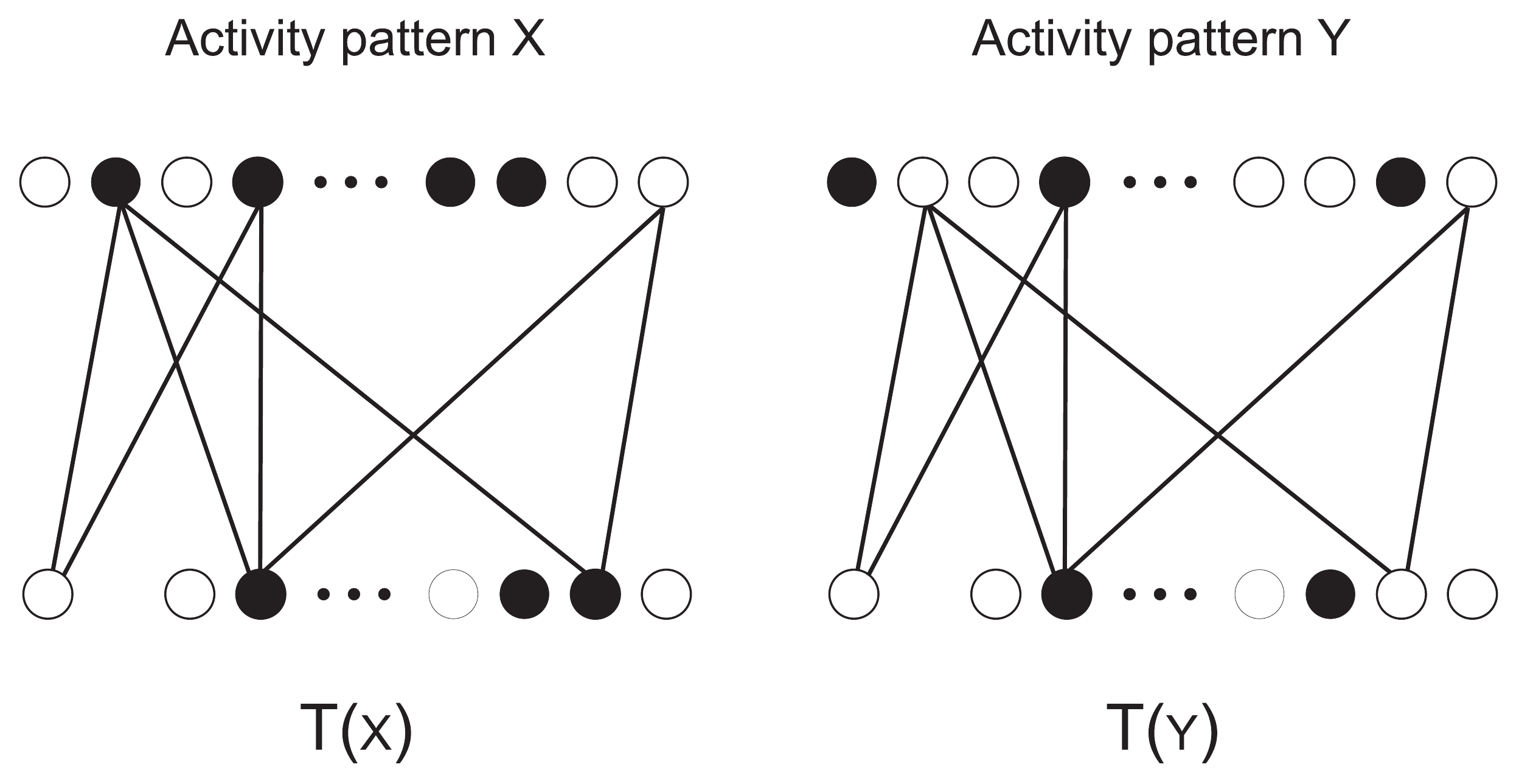} %
\caption{Transformation of inputs in a neuronal network. The input pattern is a binary vector that represents activity in input neurons (top circles) within a short time window. Ones in this vector correspond to spiking neurons (filled circles), zeros - to the neurons that do not spike (empty circles). Input neurons are connected to the neurons in the output network (bottom circles). Each input pattern makes some of the output neurons spike. Given the distance between the two input patterns X and Y how close are the output binary patterns T(X) and T(Y)?}
\vspace{0.15in}
\label{network_transformation}
\end{figure}

The current study was motivated by our recent analysis of a basic characteristic of input-to-output transformations in a hippocampal network. In that study input vectors had realistic dimensions of the order of tens of thousands \cite{Oly12}. In those vectors each vector component represented a neuron. If a neuron spiked during the time window considered the corresponding vector component was equal to one, otherwise it was equal to zero. Output vectors represented neuronal responses -- spikes or no spikes -- to inputs. The goal was to determine how distances between pairs of input binary vectors transformed into the distances between the pairs of the corresponding output binary vectors (Fig.\ref{network_transformation}).  On one side, this transformation of distances is a fundamental mathematical characteristic of a map, performed by a network. On the other side, it allows one, for example, to contrast normal and abnormal information processing in neuronal networks.  Indeed, intuitively, if an input pattern makes a target neuron spike then the "healthy" target neuron should also spike in response to similar patterns - otherwise, neurons would be too sensitive to noise.  At the same time neurons should discriminate between sufficiently different input patterns and spike selectively.
To proceed with computationally demanding studies of how particular neuronal properties affect  information processing we needed a deeper understanding of the mathematical properties of the problem.

\begin{figure} [t]
\centering
\includegraphics [scale = 0.5]{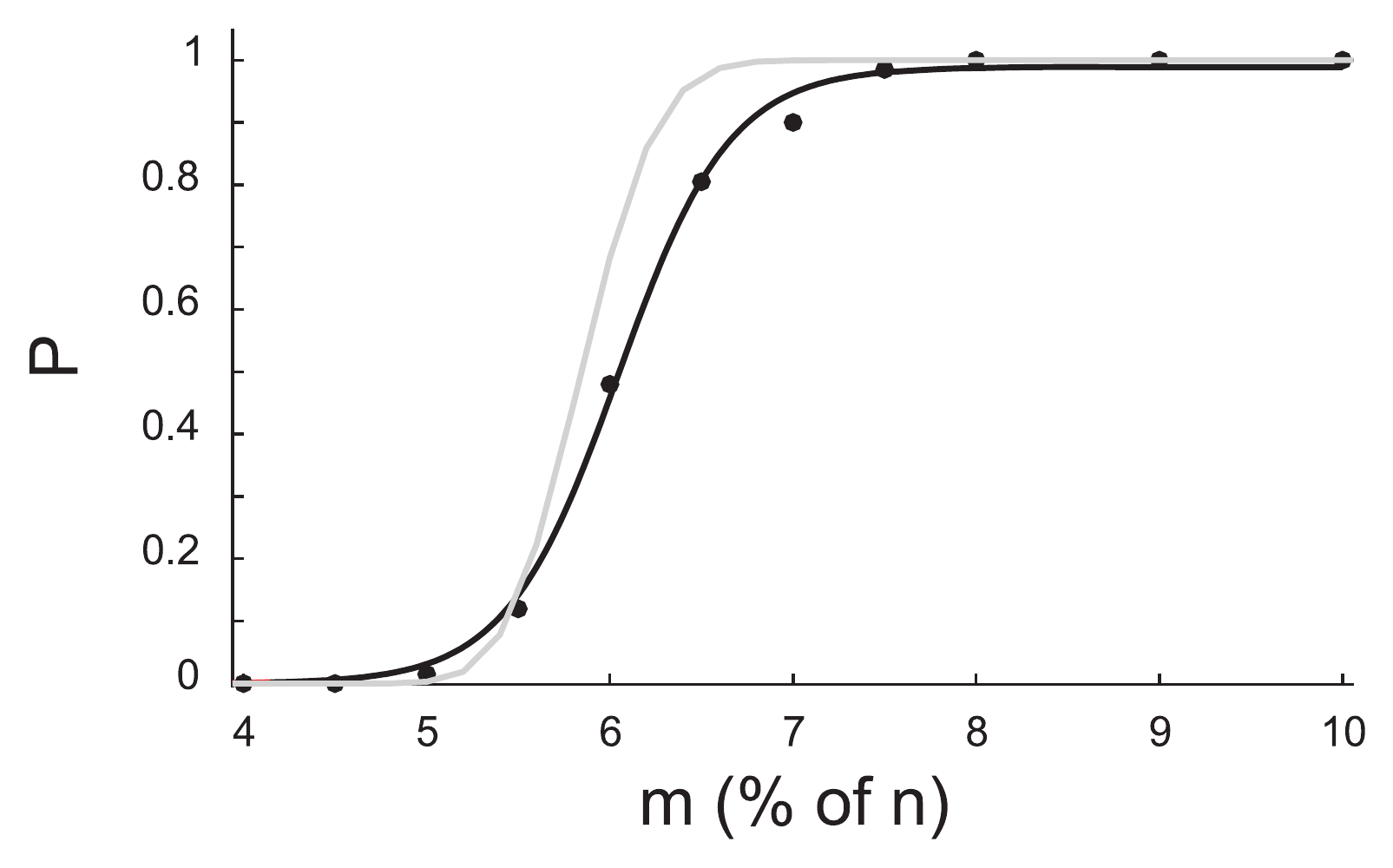}  %
\caption{Probability of spiking in a detailed neuronal model and a perceptron. Data points for the probability $P$ of spiking for the detailed neuronal model (dots) are the averages for $200$ random input patterns generated for various values of $m$, which is the number of ones in the binary vector of length $n$. The data points are fitted with a Boltzmann function (black curve). For the perceptron the (gray) curve was calculated exactly as a sum of probabilities that a  weighted combination of the input pattern components exceeds the threshold. Each probability obeyed the hypergeometric distribution with parameters $n$, $m$, and $k$.  The perceptron threshold value $258$ was chosen to match the perceptron and detailed neuronal model curves for $m=5.5\%$ which is a feasible number of ones in the physiological case considered; see discussion in \cite{Oly12}. Input binary vectors had dimension $n=28009$. The perceptron and the detailed neuronal model were connected with randomly chosen $4407$ out of $28009$ input neurons. The neuronal model is from \cite{Jar05, Oly12}.}
\vspace{0.2in}
\label{Jarsky_perceptron}
\end{figure}

That motivated us to analyze input-to-output transformations in networks of simple model neurons, perceptrons. Perceptrons  \cite {McP43, Ros58, Ros62}  continue to be used in theoretical analysis of  information processing in real neurons (see for example \cite{Bru04, Its08, LM08, Val12}). When learning is not considered, as in this study, the difference between input-to-output transformations in perceptrons and real neurons for certain ranges of inputs can be small. As Figure~\ref{Jarsky_perceptron} shows,  the input-to-output characteristics of the perceptron with an appropriate threshold value, and a detailed neuronal model comprising several thousand nonlinear differential equations are close, especially when approximately 5.5\% of the input neurons spike (5.5\%  is a typical level of activity in the input network considered; see discussion in \cite{Oly12}). Both neuronal models in this figure had the same connectivity with the input network.

 Mathematically, perceptron is a linear threshold function which is a composition of a weighted sum of the input vector components with a threshold function.  Many mathematical properties of perceptrons have been established decades ago (see, for example, \cite {Cov65}). Yet they still attract attention of mathematicians \cite{Gol06, Kli05, CB04}. Among various directions of research the following two are related to the present study.  One is the analysis of the generalization error of perceptrons (\cite{Kli05, Fe98}). The other is the analysis of kernels (\cite{HSS08}). Both directions has been developing in the context of pattern discrimination.  The problem of the input-to-output transformation of the distances between inputs and outputs of perceptrons hasn't been solved as far as we know. The only results that we are aware of are based on approximations and computer simulations \cite {Ber00, Yan13, Ze09}.  

Using a combinatorial approach we got exact formulas for the transformation of distances between pairs of inputs by linear threshold functions with binary coefficients; as we discuss below, binary coefficients under the circumstances considered  is a feasible approximation. Numerical analysis of those formulas led us to conclusions that are potentially interesting to neurobiologists and could guide simulations of complex neuronal networks. 

The outline of the paper is as follows. In section~2, we introduce basic definitions and notations and solve the problem in a simple case when the coefficients of the linear threshold function are equal to one. In section~3, we expand those results to a general case in which some weights may be equal to zero. In section~4, we apply the obtained formulas to demonstrate that connectivity and excitability of the neuron together optimize its ability to discriminate between similar and distinct inputs. 

\section{Definitions, notations, and auxiliary results}
We consider binary vectors  $x\in {V^n=\{0,1\}^n}$, and linear threshold functions in $V^n$. A linear threshold function $L$ is determined by a pair $(w,\theta)$, $w\in \mathbb {R}^n$, $\theta \in \mathbb {R}$. By definition, $L(x)=1$, if  $\langle w,x\rangle >\theta$ and $L(x)=0$ otherwise. In other words, $L$ defines a bipartition $V^n=V_+^n\cup V_-^n$ with $V_+^n $ consisting of vectors above the hyper-plane $\langle w,x\rangle=\theta$, and $V_-^n $ consisting of vectors at or below the hyper-plane;  $V_+^n $ is the support, $supp\;(L)$, of $L$.  In what follows we assume that all the components of $w$ are binary, $w \in V^n$. We refer to such functions $L$ as binary linear threshold functions.

The Hamming distance between $x,y \in V^n$ is $H(x,y)=\langle x-y, x-y \rangle$, $0 \leq H(x,y) \leq n$.  The number of ones in $x$ is the Hamming weight $|x|$ of $x$, $|x|=\sum_{i=1}^n x_i=\langle x,x\rangle$.  $V_m^n$, $0\leq m\leq n$, is the subset of $V^n$ vectors with Hamming weight $m$~:  $\displaystyle{V_m^n=\{x\in V^n, |x|=m\}}$. 

The following simple property of the Hamming distance is of special importance for this study: the expected Hamming distance between network outputs is the sum of the expected Hamming distances between individual neurons. Indeed, consider $N$ output neurons (Fig.~\ref{network_transformation}).  Let $P_\omega$ be the probability of an input pair $\omega$. Then the Hamming distance between the corresponding outputs of the $i-$th neuron, $H^i(\omega)$, $i=1,\cdots, N$, is a random variable with the expected value $$E(H^i)=\sum_{\omega\in\Omega}P_\omega H^i(\omega).$$ 
Consider now the Hamming distance between the outputs of the whole network $H_N=\sum_{i=1}^NH^i$. By definition of the expected value, 
\begin{equation}\label{networkH}
E(H_N)=\sum_{i=1}^NE(H^i). 
\end{equation}

In the case of identical neurons with the same number of connections all $E(H^i)$ are equal, $E(H^i)= \hat H$, and $\hat H_N \equiv E(H_N)=N\hat H$. A trivial generalization holds for networks that consist of several categories of identical neurons. In that case the expected Hamming distance between the outputs of the network is equal to the sum of the products of the expected Hamming distances for individual neurons from the categories and the numbers of neurons in those categories.

\begin{figure} [t]\label{example1-3}
\centering
\includegraphics[scale = 0.7]  {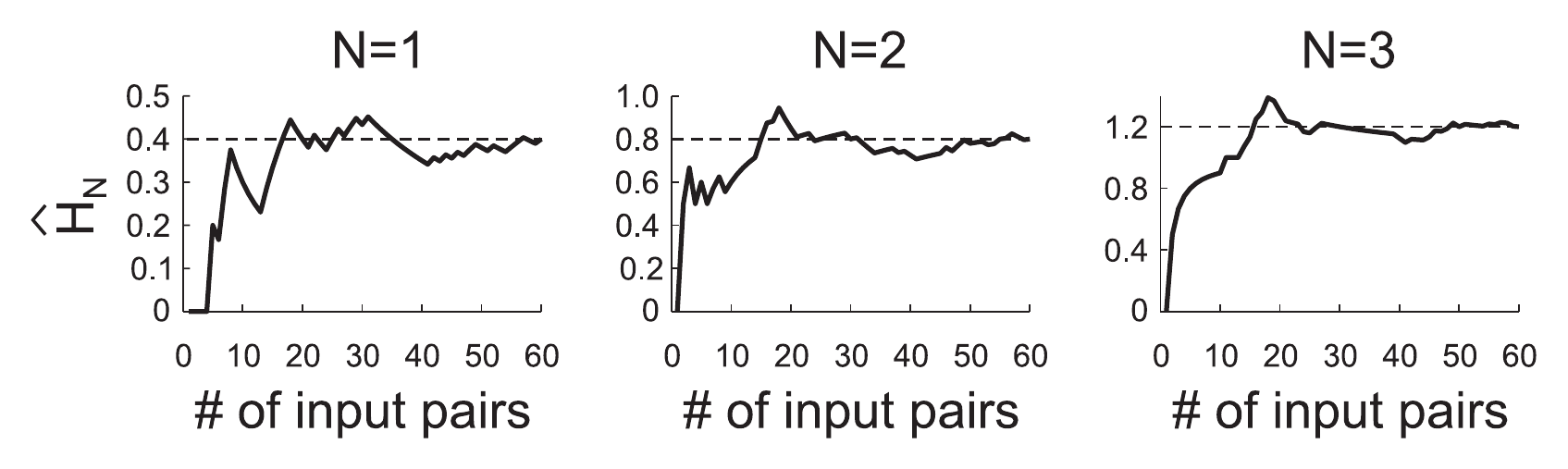} %
\caption{Mean Hamming distance between outputs of networks. The output networks consisted of one (left), two (middle), and three (right) neurons. The first neuron was connected with neurons number 1, 3, 4 of the input network. For the second and third neurons the connections were the neurons number 1, 2, 3, and number 2, 3, 5 respectively. The curves represent the average Hamming distance $\hat H_N$ across pairs of 5-dimensional input binary vectors with two ones in each vector and Hamming distance 2 between them. The pairs of inputs were selected in random order, the same for the all three graphs. The number of all such pairs is equal to 60. For that number of pairs, $\hat H_N$ was equal to exact values $0.4$, $0.8$, and $1.2$ respectively (dashed lines).}
\vspace{0.2in}
\end{figure}

\vspace{0.15in}
Figure~\ref{example1-3} illustrates formula (\ref{networkH}) for the case of five input neurons and several identical neurons in the output network.  For that particular perceptron $\hat H=0.4$; see (\ref{Hmean}) below. The figure shows convergence of  $\hat H_N$ to $0.4$, $0.8$ and $1.2$ for one, two, and three output neurons with the increase of the number of sampled pairs of inputs.

The above computations also illustrate the lack of independence of output neurons. In the case of independence, the probabilities of the Hamming distance $H_N$  between the network outputs would obey the binomial distribution with the parameters $N=3$ and $p=0.4$. The probabilities for $H_N=0$, $1$, $2$, and $3$ would be $0.216$, $0.432$, $0.288$, and $0.064$ respectively. However, computations give the values $0.133$, $0.600$, $0.200$, and $0.067$.

\vspace{0.15in}
Below, we also focus on another entity -- the probability of $y \in supp\;(L)$ conditional on $x \in supp\;(L)$ and $H(x,y)=d$. The probability characterizes the sensitivity of a linear threshold function $L$ to differences in inputs.

 To calculate the probability we introduce the function $f(n,m,m',d)$ equal to the number of the pairs $(x,y)$, $x\in V_m^n$, $y \in V_{m'} ^n$, at distance $d$ from each other. A direct counting of appropriate pairs shows that

\begin{equation} \label{f}
\begin{array}{l}
f(n,m,m',d) = {\displaystyle{\binom{n}{m} \binom{m}{\langle x,y\rangle}  \binom{n-m}{m'-\langle x,y\rangle)}}}%
= {\displaystyle{\binom{n}{m} \binom{m}{0.5(m+m' - d)}  \binom{n-m}{0.5(m' - m + d)}}}.
\end{array}
\end{equation}

\noindent
In the above formula, $\langle x,y\rangle = 0.5(m+m' - d)$ and $m'-\langle x,y\rangle = 0.5(m' - m + d)$ because of the assumption $H(x,y)=\langle x-y, x-y\rangle=d$.

\vspace{0.15in}
In what follows we consider pairs of $x, y \in V_m^n$ such that $H(x,y)=d$.  The probability mass function of $H$ can be easily obtained using $f(n,m,m',d)$. 

\noindent
\begin{propo} 
For $x, y \in V_m^n$  
 
\begin{equation}\label{formula_propo1_joint}
Prob(H(x,y)=d)= \frac{\displaystyle{\binom{m}{\frac{d}{2}}\binom{n-m}{\frac{d}{2}}}}{\displaystyle{\binom{n}{m}}}.
 \end{equation}

\label{propo1_joint}
\end{propo}
\begin{pf} Formula (\ref{formula_propo1_joint}) is the ratio of the number of combinations of $x, y \in V_m^n$ such that $H(x, y)=d$, and the total number of combinations of $x, y \in V_m^n$.  The first number is given by $f(n,m,m,d)$. The second number is equal to ${\binom{n}{m}}^2$. Obvious simplifications  lead to the result. ~$\square$
\end{pf}

{\it Example.} Let $n=3$, $m=m'=2$.  Direct counting shows that the probability of $H(x,y)=0$ equals $1/3$ (three combinations of $(x,y)$ out of nine such that $H(x,y)=0$), and the probability of $H(x,y)=2$ equals $2/3$  (six combinations of $(x,y)$ out of nine such that $H(x,y)=2$). These numbers are in accord with (\ref{formula_propo1_joint}). ~$\square$

\vspace{0.15in}
Function $f(n,m,m',d)$ can be also interpreted as the number of ways of putting $n$  pairs of vector components  $(x_i,y_i)$,  $i=1,\ldots, n$,  into 4 distinct categories: $(1,1)$, $(1,0)$, $(0,1)$, and $(0,0)$. 
Indeed, the expanding of the binomial coefficients in (\ref{f}) shows that $f(n,m,m',d)$  is the multinomial coefficient 
$$f(n,m,m',d) = \displaystyle\frac{n!}{\langle x,y\rangle!(m-\langle x,y\rangle)!(m'-\langle x,y\rangle)!(n-m-m'+\langle x,y\rangle)!}.$$ 

\noindent
Other forms of (\ref{f}) are presented in Appendix A.

\vspace{0.15in}
Another simple formula that we'll need concerns the expected Hamming distance between two vectors of Hamming weight $m$.  The Hamming distance has a binomial distribution. The probability that the Hamming distance between a component of one vector and the correspondent component of the other vector is equal to $2\frac{m}{n}\cdot(1-\frac{m}{n})$. The expected Hamming distance between the vectors is therefore equal to $2m\cdot (1-\frac{m}{n})$. The same result follows from (\ref{formula_propo1_joint}) after applying a known identity (see (\cite{Concrmath}); section 5.2)

$$\sum_{\delta=0}^{m}\binom{m}{\delta} \binom{n-m}{\delta}\delta = (n-m)\binom{n-1}{m-1},$$

\noindent
where $\delta=d/2$.

\vspace{0.15in}
The main auxiliar result is as follows. Consider $L$ with all the weights equal to one, $|w|=n$ ('uniform weighing'), so that $\langle w, x\rangle=|x|$. 

\noindent
\begin{propo} \label{propo_u}
Let $P_u=Prob(\langle w, y\rangle>\theta\;\vert\; x\in V_m^n, H(x,y)=d)$, and $|w|=n$. Then 

\begin{equation}\label{propo1_1}
P_u=\frac{\displaystyle{\sum_{m' \in D_{[\theta+1,n]}} f(n,m ,m' ,d )}} {\displaystyle{\sum_{m' \in D_{[0,n]}} f(n,m,m', d)}}, 
\end{equation}

\vspace{0.15in}
\noindent
where $D_{[a,b]}$ is the set of all $m'$ from $[a,b]$ such that: 1) $m+m'-d$ is an even number, and 2) $ max({d-m, m-d}) \leq m' \leq min({m+d, 2n-m-d})$. 
\end{propo}
\begin{pf}  Denominator in (\ref{propo1_1}) is the number of all possible combinations of $x$, $|x|=m$, and $y$ such that $H(x,y)=d$. Numerator is the number of those combinations that satisfy an additional condition $\langle w, y\rangle=|y|=m'>\theta$.  The conditions for $D_{[a,b]}$ are those for which all binomial  coefficients that involve $m'$  in the corresponding sums have non-negative integer coefficients. ~$\square$
\end{pf}

\begin{rmk}
Another way of proving Proposition \ref{propo_u}  is to notice that the scalar product $\langle x, y\rangle$ follows the hypergeometric distribution with parameters $n$, $m$ and $n-d$. The probability that $\langle w, y\rangle>\theta$ is then equal to the sum of probabilities of all values $\langle x, y\rangle$ such that $\langle x, y\rangle > (\theta+m-d)/2$. The latter inequality follows from the expansion of the scalar product expression for $H(x,y)$, $H(x,y)=\langle x-y, x-y\rangle$, condition $\langle y, y\rangle > \theta$, and the premises of the proposition. ~$\square$
\end{rmk}

The formula for conditional probability (\ref{propo1_1}) in particular holds for $x \in V_m^n \cap V_+^n$ that is when $|x|=m >\theta$. It therefore gives the probability of selecting a vector of weight $m$ from $supp\;(L)$ at distance $d$ from a given vector of weight $m$ from $supp\;(L)$.  See further analysis of formula (\ref{propo1_1}) in Appendix B. 

To evaluate these and other formulas we used Matlab (MathWorks, Natick, MA) and PC with a 1.5 GHz processor and 2.5 Gb memory. To preserve accuracy, we made calculations with all the digits utilizing a publicly available Matlab package VPI by John D'Errico \\ 
(http://www.mathworks.com/matlabcentral/fileexchange/22725). 

\section{Arbitrary binary weighing}
Here we  generalize the results of the previous section to the case when some weights of the linear threshold function are equal to one while the others are equal to zero ('arbitrary weighing'). In neuronal models zero weights correspond to "silent", ineffective connections between neurons. The role of such neurons in neuronal information processing is intensely studied; see for example \cite{Bru04, Bru13}.

Below we assume that the non-zero weights are the first $k$ weights of $L$: $w_i=1$, $i \leq k$, $w_i=0$, $i>k$.  There is no loss of generality in this assumption since all possible binary inputs are considered. 
Let $P_k$ be a projector to the first $k$ coordinates, $P_kx=(x_1, x_2, \ldots,x_k, 0, \ldots, 0)$. Denote  $\mu = \langle P_k x,P_k x\rangle$, $\mu'=\langle P_k y,P_k y\rangle$, and  $\delta =H(P_k x,P_k y)$.  The following proposition determines the probability that $y\in supp\;(L)$, provided $x \in supp\;(L)$, $|x|=m$ and $H(x,y)=d$.  

\begin{propo}\label{propo2} Let $x \in  V_m^n$, $y \in  V^n$, $w\in V^n$, $|w|<m$ and $P_a = Prob(\langle w, y\rangle>\theta\;\vert\;  \langle w, x\rangle>\theta, H(x,y)=d)$. Then

\begin{equation}\label{arb}
P_a = \frac{\displaystyle{\sum_{\delta =0}^{min(d,k)} \sum_{\mu = \lfloor \theta \rfloor +1}^{min(m,k)} \sum_{m'=0}^n  \sum_{\mu' \in Q_{[\theta +1,m']}} f(k,\mu ,\mu' ,\delta ) f(n-k,m-\mu ,m'-\mu' , d-\delta )}} 
{\displaystyle{\sum_{\delta =0}^{min(d,k)} \sum_{\mu = \lfloor \theta \rfloor + 1}^{min(m,k)} \sum_{m'=0}^n  \sum_{\mu' \in Q_{[0,m']}} f(k,\mu ,\mu' ,\delta ) f(n-k,m-\mu ,m'-\mu' , d-\delta )}}, 
\end{equation}

\noindent
where $f$ is defined by formula (\ref{f}), $\lfloor \theta \rfloor$ is the largest integer not greater than $\theta$, and $Q_{[a,b]}$ is the set of $\mu'$ from $[a,b]$ for which all binomial  coefficients that involve $\mu'$  in the corresponding sum have non-negative integer coefficients.
\end{propo}

\begin{pf}
To derive (\ref{arb}) consider first $k$ components of vectors $x$ and $y$ separately from the rest $n-k$ components.  For the first $k$ components we can assume that all the weights of $L$ are equal to one. Therefore the number of all the combinations of $P_kx$ and $P_ky$ such that $\langle P_kx, P_kx \rangle = \mu$, $\langle P_ky, P_ky \rangle = \mu'$ and $H(P_kx, P_ky)=\delta$ is equal to $f(k,\mu ,\mu' ,\delta )$; cf. (\ref{f}). Each of these combinations is multiplied by the number of possible combinations of the rest $n-k$ components.  For the latter combinations the weights of $L$  also can be considered uniform (all equal to zero). The number of combinations is also given by function $f$ from (\ref{f}) with appropriate arguments. Subsets $Q_{[a,b]}$  are natural generalizations of subsets $D_{[a,b]}$ from Proposition \ref{propo_u}. They specify values of $\mu'$ from intervals $[a,b]$ such that all binomial coefficients involving $\mu'$ in the corresponding sums have non-negative integer lower indexes. ~$\square$
\end{pf}

Note that when $k=n$ formula (\ref{arb}) reduces to formula (\ref{propo1_1}). Indeed, $m=\mu$, $m'=\mu'$, and $\delta$ takes only one value $\delta=d$.  Accordingly, $f(n-k,m-\mu ,m'-\mu' , d-\delta )=1$.

\begin{figure} [t]
\centering
\includegraphics[scale = 0.8] {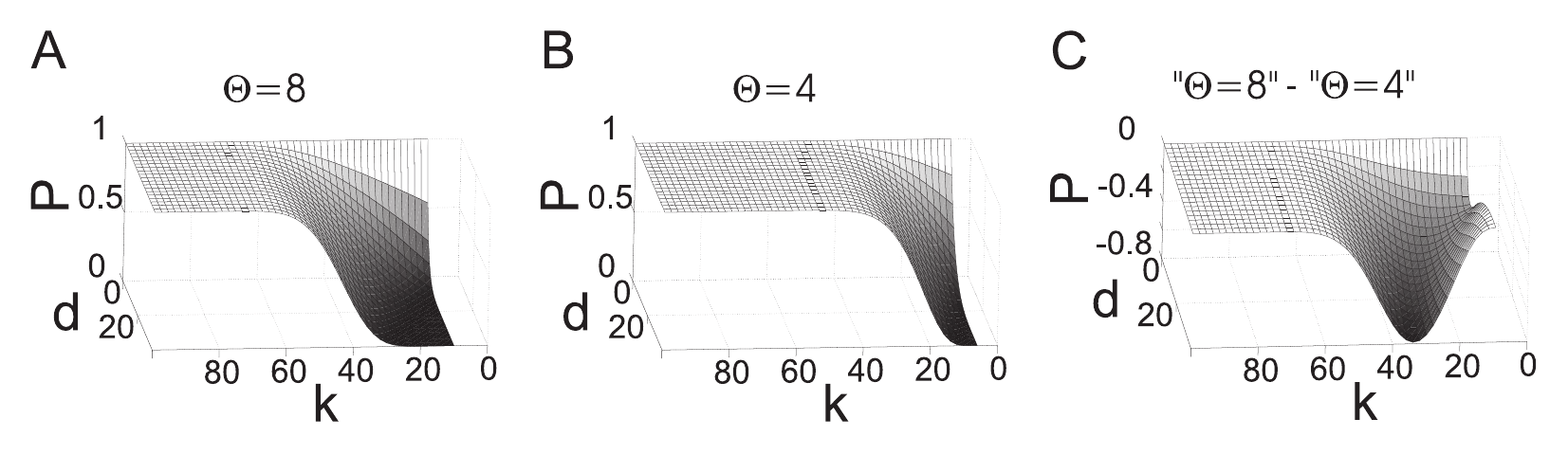} 
\caption{ Probability that a vector is in support of a binary linear threshold function. Probability $P$, conditional on the distance $d$ from another vector from support as a function of $d$ and the number of non-zero weights $k$ is shown for two different values of threshold $\theta$.  The probabilities for $\theta=8$ (A) and $\theta=4$ (B) have qualitatively the same dependency from $d$ and $k$. However, for certain values of $k$ the difference between the probabilities is very large  (C).  $n=100$, $m=m'=20$.}
\vspace{0.2in}
\label{fig_Parb}
\end{figure}

Formula (\ref{arb}) gets simpler in an important case $m'=m$, i.e. when $|x|=|y|$; see Fig.~\ref{fig_Parb}. This case is important since in applications inputs often have similar or equal Hamming weights (cf.~\cite{Buz02}).

\begin{equation} \label{arb_m}
P_a= \frac{\displaystyle{\sum_{\delta =0}^{min(d,k)} \sum_{\mu =\lfloor \theta \rfloor +1}^{min(m,k)}  \sum_{\mu' \in Q_{[\theta +1,m]}} f(k,\mu ,\mu' ,\delta ) f(n-k,m-\mu ,m-\mu' , d-\delta )}} 
{\displaystyle{\sum_{\delta =0}^{min(d,k)} \sum_{\mu = \lfloor \theta \rfloor + 1}^{min(m,k)} \sum_{\mu' \in Q_{[0,m]}} f(k,\mu ,\mu' ,\delta ) f(n-k,m-\mu ,m-\mu' , d-\delta )}}.
\end{equation}

\vspace{0.15in}
The following corollary of Proposition \ref{propo2} determines the probability that two binary inputs $x$ and $y$ from $supp\;(L)$ are at Hamming distance $d$ from each other.

\begin{corol}
Let $x, y \in V_m^n \cap V_{+}$ and $k \leq n$. Then 
 
\begin{equation}\label{distances_in_support}
Prob(H(x,y)=d)= \frac{\displaystyle{\sum_{\delta =0}^{min(d,k)} \sum_{\mu = \lfloor \theta \rfloor +1}^{min(m,k)}  \sum_{\mu' \in Q_{[\theta +1,m']}} f(k,\mu ,\mu' ,\delta ) f(n-k,m-\mu ,m-\mu' , d-\delta )}} 
{\displaystyle{\left( \sum_{\mu = \lfloor \theta \rfloor + 1}^{min(m,k)}{\binom{m}{\mu}{\binom{n-m}{k-\mu}}} \right)^2}}. 
\end{equation}
\end{corol} 

\begin{pf}
The numerator in (\ref{distances_in_support})  is the same as in formula (\ref{arb_m}) and the denominator is the total number of combinations of $x, y \in V_m^n \cap V_{+}$ for the case $k \le n$. ~$\square$
\end{pf}

{\it Example.} Let $n=5$, $k=3$, $\theta=0$, $m=m'=2$, $d=2$.  Direct counting shows that there are 54 combinations of $(x,y)$ such that $\langle x,x\rangle=2$, $\langle y,y\rangle=2$, $H(x,y)=2$, and $\langle P_k x,P_k x\rangle > 0$. In (\ref{arb_m}), 54 is the value of denominator and 48 is the value of numerator.  The probability of interest is therefore equal to $48/54=0.89$.

For the same function $L$ with $n=5$, $k=3$, and $\theta=0$, formula (\ref{distances_in_support}) gives the following probabilities of distances between $x, y \in V_2^5 \cap V_{+}^5$: $Prob (H(x,y)=0)=0.11$, $Prob (H(x,y)=2)=0.59$, $Prob (H(x,y)=4)=0.30$.

\vspace{0.15in}
Using the above results we now obtain a formula for the expected Hamming distance $\hat H$ between $L(x)$, and $L(y)$ provided  $x$ and $y$ have the same Hamming weight $m$ and are at Hamming distance $d$ from each other.

\begin{propo}  Let $x, y \in V_m^n$, $H(x,y)=d$ and  $L$ is a linear threshold function with the weights $w\in V_k^n$ and threshold $\theta$. Then $\hat H$, the expected Hamming distance between $L(x)$ and $L(y)$, can be calculated using the formula

\begin{eqnarray}
\hat H = Prob(L(y)=1|L(x)=1, H(x,y)=d) \cdot Prob(L(x)=1|H(x,y)=d)\nonumber\\
 + Prob(L(y)=0|L(x)=0, H(x,y)=d) \cdot Prob(L(x)=0|H(x,y)=d),
\label{Hmean}
\end{eqnarray}

\noindent
where
$$
\begin{array}{l}
Prob(L(y)=1|L(x)=1, H(x,y)=d)\\
\quad =\frac{\displaystyle{\sum_{\delta =0}^{min(d,k)} \sum_{\mu =\lfloor \theta \rfloor +1}^{min(m,k)}  \sum_{\mu' \in Q_{[\theta +1,m]}} f(k,\mu ,\mu' ,\delta ) f(n-k,m-\mu ,m-\mu' , d-\delta )}} 
{\displaystyle{\sum_{\delta =0}^{min(d,k)} \sum_{\mu = \lfloor \theta \rfloor + 1}^{min(m,k)} \sum_{\mu' \in Q_{[0,m]}} f(k,\mu ,\mu' ,\delta ) f(n-k,m-\mu ,m-\mu' , d-\delta )}},\\
\\
Prob(L(y)=0|L(x)=0, H(x,y)=d) \\
\quad = \frac{\displaystyle{\sum_{\delta =0}^{min(d,k)} \sum_{\mu =0}^{\lfloor \theta \rfloor}  \sum_{\mu' \in Q_{[0,\theta ]}} f(k,\mu ,\mu' ,\delta ) f(n-k,m-\mu ,m-\mu' , d-\delta )}} 
{\displaystyle{\sum_{\delta =0}^{min(d,k)} \sum_{\mu =0}^{\lfloor \theta \rfloor} \sum_{\mu' \in Q_{[0,m]}} f(k,\mu ,\mu' ,\delta ) f(n-k,m-\mu ,m-\mu' , d-\delta )}},\\
Prob(L(x)=1|H(x,y)=d) =\frac{\displaystyle\sum_{\mu = \lfloor \theta \rfloor +1}^{min(m,k)} \binom{k}{\mu} \binom{n-k}{m-\mu}}{\displaystyle{\binom{n}{m}}}, \\
Prob(L(x)=0|H(x,y)=d) = 1 - Prob(L(x)=1|H(x,y)=d),  and \\
\\
\mu=P_kx, \mu'=P_ky.
\end{array}
$$

\end{propo}
\begin{pf}
The formula for $Prob(L(y)=1|L(x)=1, H(x,y)=d)\equiv P_a$  was defined earlier in (\ref{arb_m}). The formula for    $Prob(L(y)=0|L(x)=0, H(x,y)=d)$ is similar. The difference is that the sums are now taken across the values of $\mu$ and $\mu'$ less or equal $\theta$ since $L(x)$ and $L(y)$ should be equal to zero. The formula for $Prob(L(x)=1|H(x,y)=d)$ is obtained by straightforward counting $x$ for which $\mu>\theta$. $~\square$
\end{pf}

\vspace{0.15in}
Note that in the case of uniform weighing, ($|w|=n$), $Prob(H(L(x),L(y))=0)\equiv 1$. If $m>\theta$ then $H(L(x),L(y))=0$ since $L(x)=1$ and $L(y)=1$. If $m \leq \theta$ then $H(L(x),L(y))=0$ since $L(x)=0$ and $L(y)=0$. 

\begin{figure} [t]
\centering
\includegraphics[scale = 0.7] {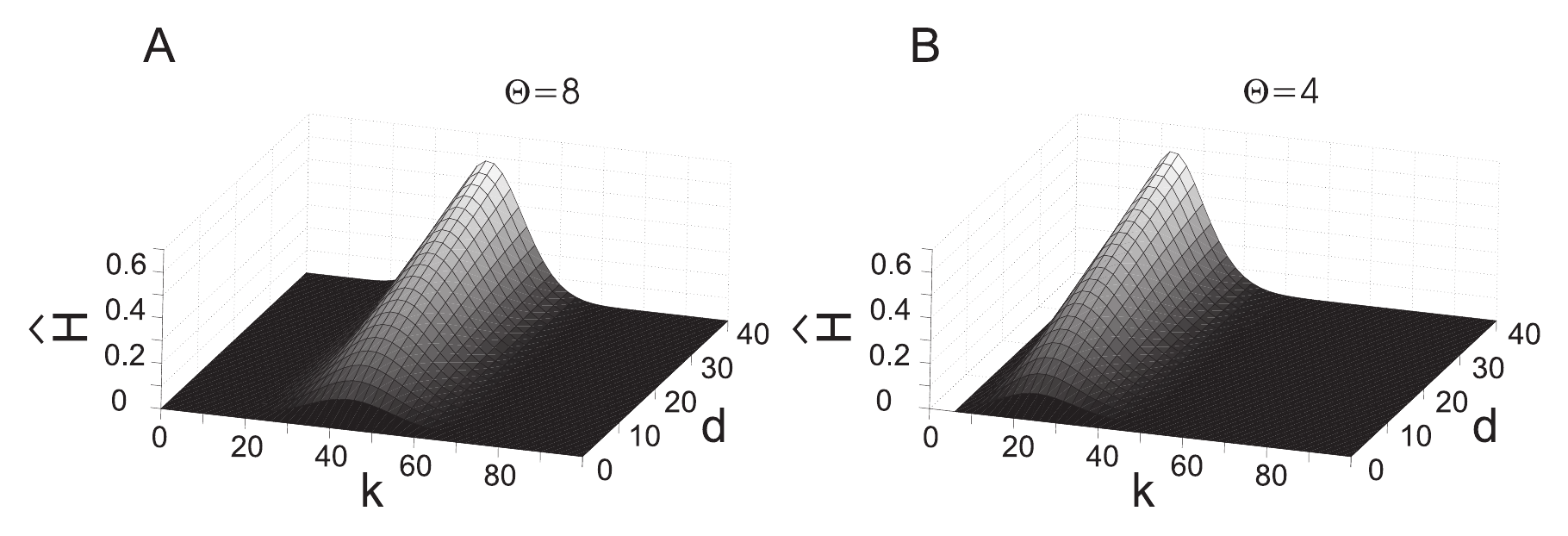} 
\caption{Expected Hamming distance $\hat H$ between images $L(x)$, $L(y)$ of a binary linear threshold function $L$  as a function of the Hamming distance $d$ between preimages $x$, $y$ and the number of non-zero weights $k$. $n=100$, $m=m'=20$, $\theta=8$.}
\vspace{0.2in}
\label{fig_Hmean}
\end{figure}

\vspace{0.15in}
Figure~\ref{fig_Hmean} illustrates how the decrease of threshold $\theta$ changes how a perceptron transforms its inputs. In accord with intuition, small distances $d$ between $x$ and $y$ are transformed into small expected distances $\hat H=H(L(x),L(y))$ between the corresponding outputs for the both considered threshold values, $\theta=8$ and $\theta=4$. However, greater values of $d$ are transformed to large values of $H(L(x),L(y))$ only for particular values of connectivity parameter $k$ dependent on $\theta$. 	Namely, for $\theta=8$ the greatest values of $\hat H$ are achieved for about two times greater values of $k$ compared to the case of $\theta=4$.

\section{Application}
In this section we  use formula (\ref{arb_m}) to explore information processing in the hippocampal field CA1.  The  hippocampus is a brain structure that is critically implicated in learning and memory (\cite{And06}).  CA1 neurons receive inputs from the neurons of another field, CA3, of the hippocampus. Neurons of CA1 produce the output of the hippocampus (\cite{Ama06}) . The anatomy of  hippocampal connectivity and excitability of hippocampal neurons are well studied (\cite{Ama06}). However,  little is known how their interplay effects the information processing in CA1. For example, an influential study (\cite{Tre04}) mostly concerns about connectivity.

For our analysis we used the data represented in Figure~\ref{fig_Parb}.  For the Hamming distance $d$ between spiking patterns in CA3 we used two values.  One value, $d=32$ was equal to the expected distance between a pair of randomly selected binary patterns with 20 ones out of 100 (cf. section 2); note that $d=32$ is 80\% out of the maximal $d=40$.  The other value, $d=4$, was equal to 10\% of the maximal distance between a pair of patterns.  

\begin{figure} [ht]
\centering
\includegraphics[scale = 0.6] {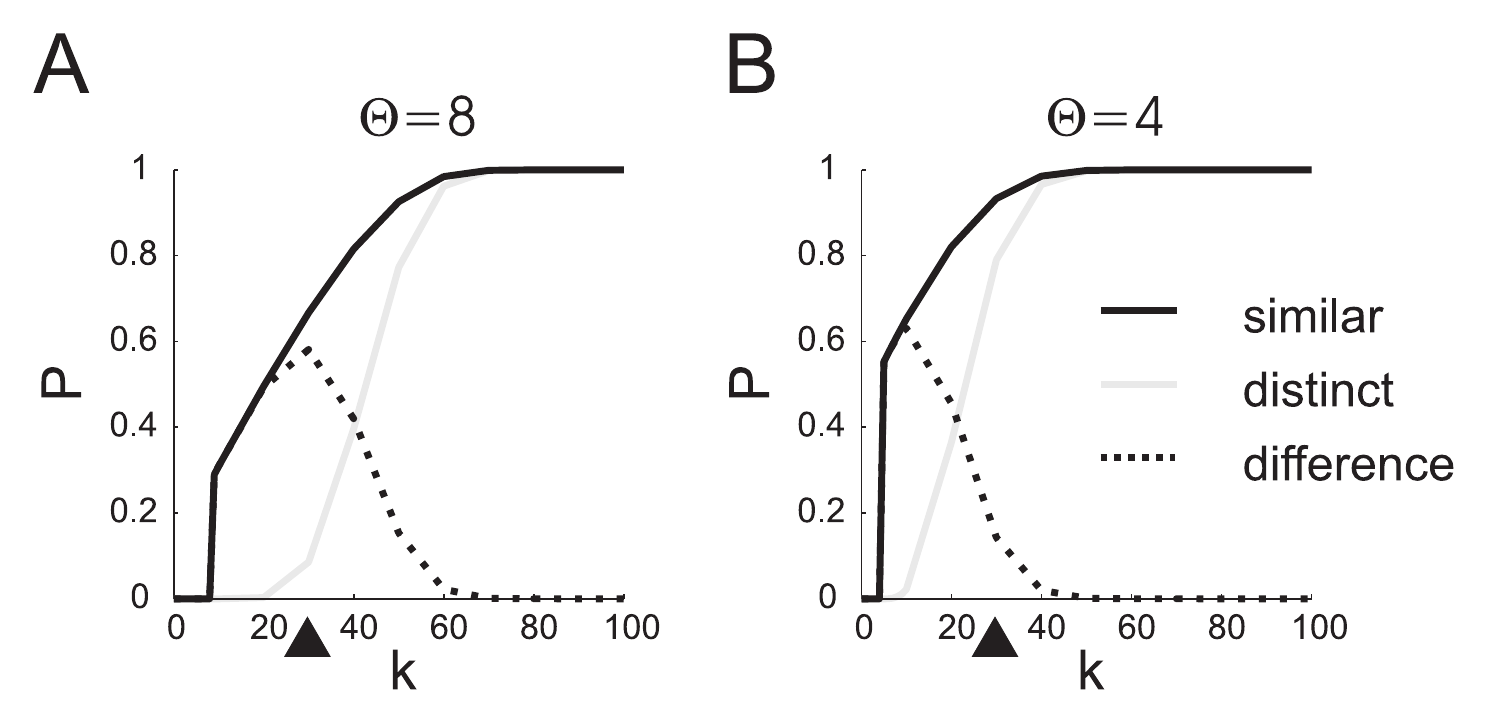} %
\caption{Probability $P$ that an input makes a neuronal model spikes conditional on the distance $d$ of that input from another input that makes the model spike. The difference between the probability for similar inputs ($d=4$) (black) and distinct inputs ($d=32$) is maximal for the number of  non-zero weights (synaptic connections) $k=30$ (black triangle) when the threshold $\theta=8$ (A). For a smaller threshold $\theta=4$, $k=30$ is no longer optimal (B). $n=100$, $m=m'=20$.}
\vspace{0.2in}
\label{fig_appl}
\end{figure}

Figure~\ref{fig_appl}A shows the probability of spiking when $y$ is similar to $x$ ($d=H(x,y)=4$; black curve) and distinct from $x$ ($d=H(x,y)=32$; gray curve). The figure also shows the difference between the probabilities (dotted curve). The difference reaches maximum for $k=30$, i.e. when a model CA1 neuron is connected with 30 CA3 neurons; the maximal value of the difference between the probabilities is equal to $0.55$. Thus when $k=30$ the model neuron discriminates between the chosen categories of similar and distinct patterns best of all.  

A decrease of the spiking threshold to $\theta=4$ made the connectivity with $k=30$ non-optimal (Fig. \ref{fig_appl}B). The two conditional probabilities for similar (black curve) and distinct (gray curve) patterns changed, along with the difference between the probabilities (dotted curve). 
As a result, the difference between the probabilities for $k=30$ became equal to $0.14$ that signifies a considerable decrease of the neuron's ability to discriminate between similar and distinct inputs.  

\section{Conclusion}
Complex information processing in the brain in certain cases can be considered as a transformation of binary vectors of spikes/no spikes of an input network within a short time window to binary vectors of spike/no spike responses of output neurons.  Such time windows are observed for example during rhythmic states of neuronal activity \cite{Buz06}. One of the basic characteristics of such a transformation is how it separates inputs. For example, does it transform close input vectors to close output vectors? A proper answer to this question would be a distribution of distances between output vectors for each distance between input vectors.  

Here we found such a distribution for the transformation of binary vectors by linear threshold functions with binary weights. The support of this distribution consists just of two elements, one and zero. The expected value of the (Hamming) distance between the values of the function for a pair of inputs is therefore equal to the probability that the function has different values on those inputs. In neuronal modeling linear threshold functions are called perceptrons \cite{Ros58}. Knowing the expected value of the Hamming distance for one perceptron is sufficient to determine the expected Hamming distance between the outputs of the network of  perceptrons; see (\ref{networkH}). Note that (\ref{networkH}) has no reference to particularities of the neuronal model. In fact the formula is applicable even to a network of biological neurons.

\begin{figure} [t]
\centering
\includegraphics[scale = 0.5] {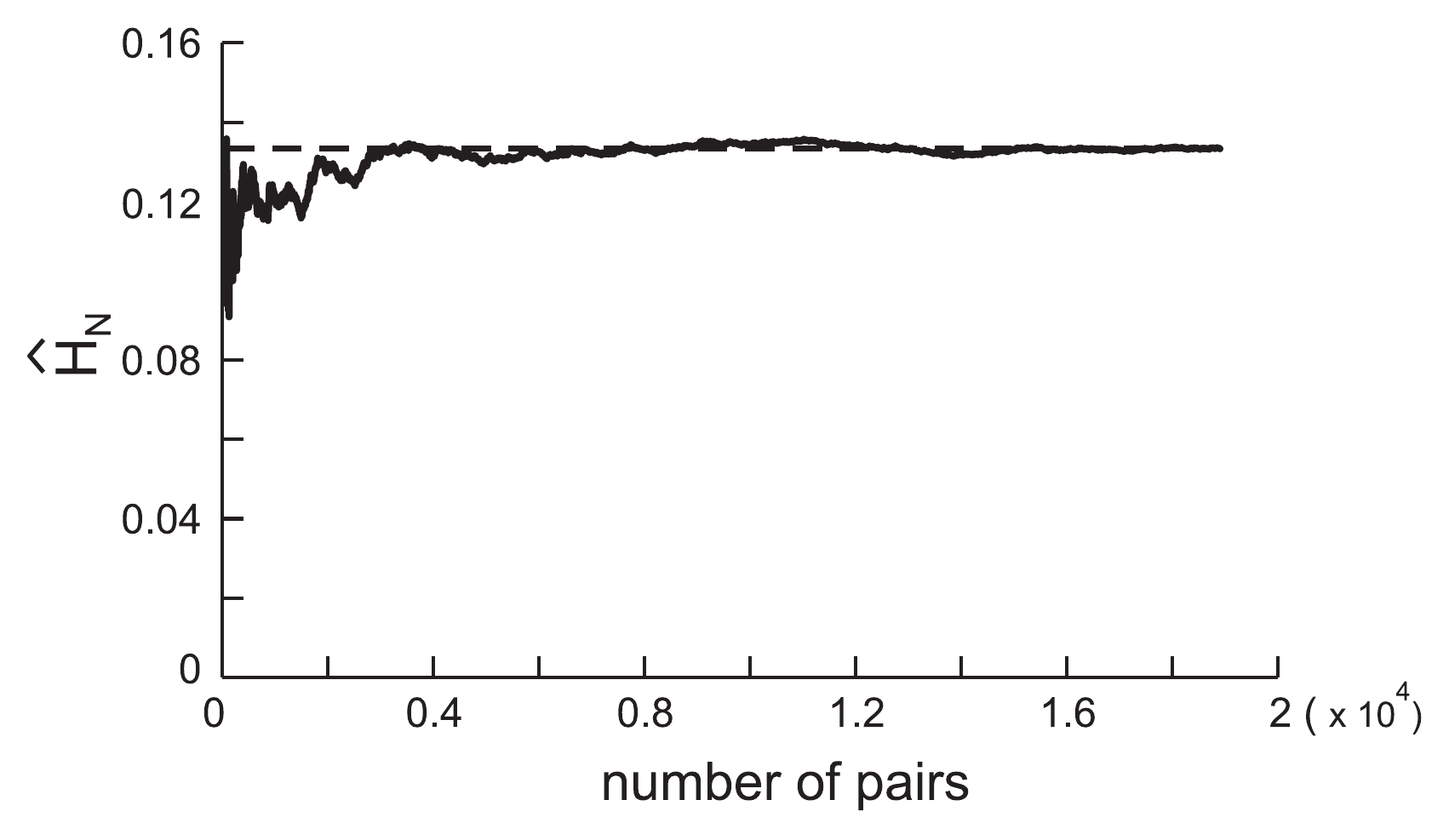} %
\caption{Expected Hamming distance for a network of  two identical perceptrons. Expected Hamming distance $\hat H_N$ was calculated for an increasing number of different pairs of input patterns -- binary vectors of dimension ten with four ones. In each pair the Hamming distance between the vectors was equal to four. With the increase of the number of the considered pairs to the maximal value of $18900$,  $\hat H_N$ converged to the exact value, $0.1333$ (dashed line).}
\vspace{0.2in}
\label{asymHmean}
\end{figure}

Obtained exact formulas for the expectations of Hamming distances can be further developed in a number of directions.  One particular question relates to asymptotic behaviors of the formulas. Consider an example of two ($N=2$) identical perceptrons with threshold $\theta=2$, each connected with three randomly chosen  input neurons out of ten ($k=3, n=10$). Consider the set of all input pattern pairs such that each pattern has exactly four active neurons ($m=m'=4$), and Hamming distance between the patterns in every pair is equal to 4 ($d=4$). According to (\ref{f}) there are $f(10,4,4,4)=18900$ of such pairs. We randomly selected pairs from this set, evaluated Hamming distance between the corresponding outputs and calculated the average distance for an increasing number of pairs. Figure~\ref{asymHmean} shows that the approximate values of $\hat H_N$ obtained for subsets of randomly chosen pairs of inputs converge to the exact value $0.133$, obtained using (\ref{networkH}) and (\ref{Hmean}). This figure also shows that $400$ pairs of inputs or approximately  $2\%$ of the total number are enough to obtain a good approximation to the exact value. An interesting question is whether there is a corresponding  asymptotic formula for (\ref{Hmean}) that would account for this result.  Another direction of subsequent research is development of perturbation formulas to extend current results to perceptrons with small random variations of weights.

The perceptron neuronal model with binary weights that we used is an extreme simplification of a biological neuron given the whole universe of the properties of the latter. However, from the perspective of input-to-output transformation of binary inputs the difference can be made relatively small by choosing a proper value of perceptron threshold (Fig.~\ref{Jarsky_perceptron}). Some support to using perceptron models comes also from recent experimental data. In particular, input-to-output transformations in real (hippocampal) neurons and networks in some cases allow for linear approximation \cite{Cash99, PB12}. The assumption of binary weights used in our study is equivalent to the assumption of equal synaptic weights. In the case of the hippocampal field CA1 the assumption is supported by the observation that excitatory synapses at different locations make similar contribution to the changes of the membrane potential in the soma of the neurons in that area \cite{Mag00, Mag03}.  The neurons in the hippocampal field CA1 play a key role not only in normal information processing. Their abnormally increased activity is a first indicator of developing schizophrenia \cite{Sch09}. In a neuronal model, increased neuronal spiking can be associated with a decreased spiking threshold. The example, considered in the Application suggests that such a decrease impairs the neuronal ability to discriminate between similar and distinct inputs. Such an impairment may be a basic element of complex cognitive symptoms of schizophrenia \cite{Sil00}.

Besides neuroscience, linear threshold functions appear in various information systems, especially computational systems \cite{Aro09}. Our formulas can be used to find optimal characteristics of such systems in terms of Hamming distances between inputs/outputs. 

\appendix
\section{}
Here we deduce a number of useful equivalent forms of (\ref{propo1_1}). First,  the coefficient$\binom{n}{m}$ in numerator and denominator can be canceled out.  Denominator  

\begin{equation} \label{denominator}
\begin{array}{l}
{\displaystyle{S=\sum_{m'=0}^{n} \binom{m}{0.5(m+m' - d)} \binom{n-m}{0.5(m' - m + d)} }}\\
 {\displaystyle{= \sum_{m'=0}^{n}\binom{m}{0.5(m-d)+0.5m'} \binom{n-m}{0.5(d-m) +0.5m')}}}
\end{array}
\end{equation}

\vspace{0.15in} 
\noindent
can be simplified using a variant of Vandermonde's identity (\cite{Concrmath}, Eqn.(5.23))

\begin{equation}\label{Vandermonde}
\sum_{k} \binom{l}{m+k} \binom{s}{n+k} = \binom{l+s}{l-m+n},
\end{equation}

\vspace{0.15in} 
\noindent
that is valid for nonnegative integer  $l$, and integer $m$, $n$.  When $m-d$, and consequently $m'$ are both even then the application of (\ref{Vandermonde}) to (\ref{denominator}) yields $S=\binom{n}{d}$.  In the case when $m-d$ is odd $m'$ is also odd, and (\ref{Vandermonde}) can be applied to 

\[S= \sum_{m'=1}^{n}\binom{m}{0.5(m-d+1)+0.5(m'-1)} \binom{n-m}{0.5(d-m+1) +0.5(m'-1))}.\]

\vspace{0.15in}
\noindent
The result is the same, $S=\binom{n}{d}$, and the formula (\ref{propo1_1}) becomes

\begin{equation}\label{propo1_2}
P_u  =\frac{\displaystyle{\sum_{m'=\theta+1}^{n} \binom{m}{0.5(m+m' - d)} \binom{n-m}{0.5(m' - m + d)}}} {\displaystyle{\binom{n}{d}}}.
\end{equation}

\vspace{0.15in}
The symmetry relation $\binom{n}{k}=\binom{n}{n-k}$ applied to $\binom{m}{0.5(m+m' - d)}$, turns formula (\ref{propo1_2}) into a sum of the probabilities of the hypergeometric distribution with parameters $n$, $d$, $m$:

\begin{equation}\label{propo1_hg}
P_u =\frac{\displaystyle{\sum_{m'=\theta+1}^{n} \binom{m}{0.5(m-m' + d)} \binom{n-m}{0.5(m' - m + d)}}} {\displaystyle{\binom{n}{d}}}.
\end{equation}
\vspace{0.2in}

\section{} 
\addtocounter{corol}{-1}
\addtocounter{figure}{-7}

\begin{figure} [t]
\centering
\includegraphics[scale = 0.6]{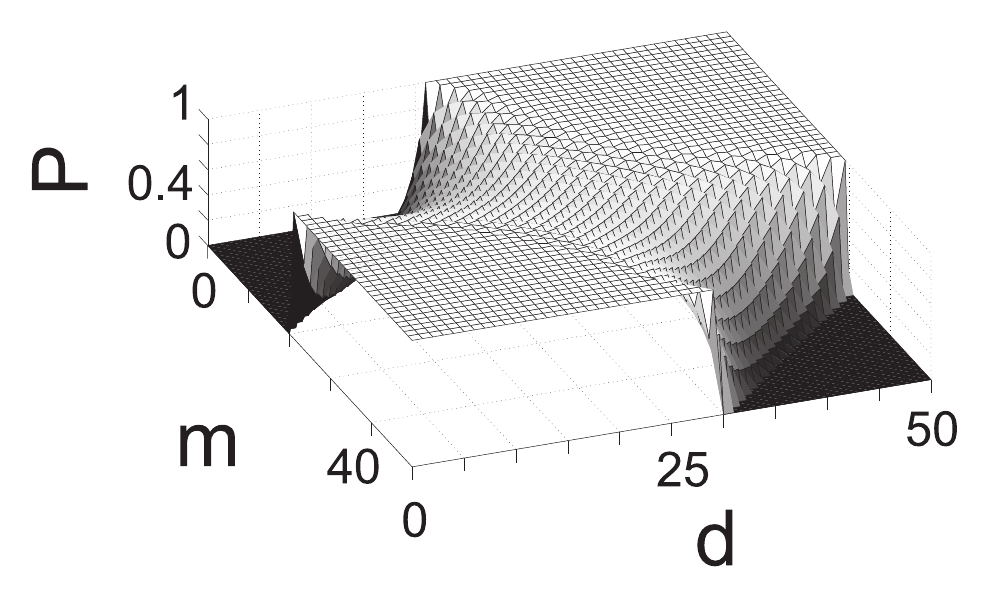} %
\caption{Probability that a binary linear threshold function $L(y)=1$ provided $H(x,y)=d$, and $|x|=m$. The probability $P$ is calculated according to (\ref{propo1_hg}). The saw-like behavior of the probability is due to the requirement that the numbers that determine binomial coefficients in (\ref{propo1_hg}) are integers.  $n=50$, $\theta=20$, $|w|=n=50$.}
\vspace{0.2in}
\label{fig_propo1}
\end{figure}

\vspace{0.15in}
This appendix contains three corollaries of Proposition \ref{propo_u} that help to reveal the properties of formula (\ref{propo1_1}); see also Fig.~\ref{fig_propo1}. The first corollary of Proposition \ref{propo_u} sets the bounds for how a pattern $y$ should be different from a pattern $x$ to belong in $supp\;(L)$. 
The bounds are formed by the values of  $d$, $m$ and $\theta$ for which the conditional probability $P_u$ from Proposition \ref{propo_u}  is equal to one.  

\begin{corol} \label{corol1}
Let $x\in V_m^n$, $y\in V^n$,  $H(x,y)=d>0$, and $|w|=n$. Then 
$$P_u = 1\quad\hbox{iff}\quad d\leq m-\theta \hbox{  or} \quad m+\theta \leq d.$$
\end{corol}

\begin{pf}
The probability in Proposition ~\ref{propo_u} is equal to one if and only if  \\
$D_{[\theta,n]}=D_{[0,n]}$, or  $max({\theta, d-m, m-d})=max({0, d-m, m-d})$.  The latter inequality holds if and only if $\,\theta \leq m-d\,$ or $\,\theta \leq d-m\,$. Rewriting those conditions as inequalities for $d$ finalizes the proof. ~$\square$
\end{pf}

According to the inequality $d\leq m-\theta$ from the corollary if $x\in supp\;(L)$, i.e. $|x|=m>\theta$, then $y$ close to $x$ is also in $supp\;(L)$.  The other inequality states that $y\in supp\;(L)$ if it is sufficiently different from $x$ regardless to whether $x\in supp\;(L)$ or not.

\begin{corol} \label{corol2}
Let $x\in V_m^n$, $y\in V^n$,  $H(x,y)=d>0$, and $|w|=n$. Then 
\[P_u = 0\quad\hbox{iff}\quad d < \theta-m\quad\hbox{or}\quad d > 2n-m-\theta.\]
\end{corol} 
\begin{pf} The probability in Proposition~\ref{propo_u} is equal to zero if and only if the set $D_{[\theta,n]}$ is empty.  Solving the corresponding inequality \[max({\theta, d-m, m-d}) >  min({n, m+d, 2n-m-d})\] yields the two possibilities stated in the corollary. ~$\square$
\end{pf}

The corollary shows that if $x\notin supp\;(L)$, i.e. $|x|=m \leq \theta$,  then $y$ close to $x$ or sufficiently different from $x$ is also not in $supp\;(L)$. Figure~\ref{fig_propo1} illustrates properties of the conditional probability  $P_u$ from (\ref{propo1_1}) and the conditions specified in the Corollaries \ref{corol1}, \ref{corol2}. 

The last corollary considered specifies the probability that a pattern $y$ has the same number of ones as a pattern $x$ provided certain Hamming distance between the patterns.  

\begin{corol} For $x\in V_m^n$ and  $y\in V^n$ such that $H(x,y)=d$ 

\[Prob(y\in V_m^n \;\vert\;  x\in V_m^n, H(x,y)=d)=
\left\{
\begin{array}{l}
\frac{\displaystyle{\binom{m}{\frac{d}{2}}\binom{n-m}{\frac{d}{2}}}}{\displaystyle{\binom{n}{d}}}\quad\hbox{if}
\quad \displaystyle{ \frac{d}{2}\leq m\leq n-\frac{d}{2}, }\\
0\;\hbox{otherwise.}
\end{array}
\right.\]
\label{corol3}
\end{corol}
\begin{pf}
The corollary follows directly from the formula (\ref{propo1_2}). In numerator, the sum reduces to only one term with $m'=m$. Inequalities in the corollary follow from the condition that the low values in the binomial coefficients are non-negative integers.  ~$\square$
\end{pf}

\begin{figure} [t]
\centering
\includegraphics[scale = 0.4]{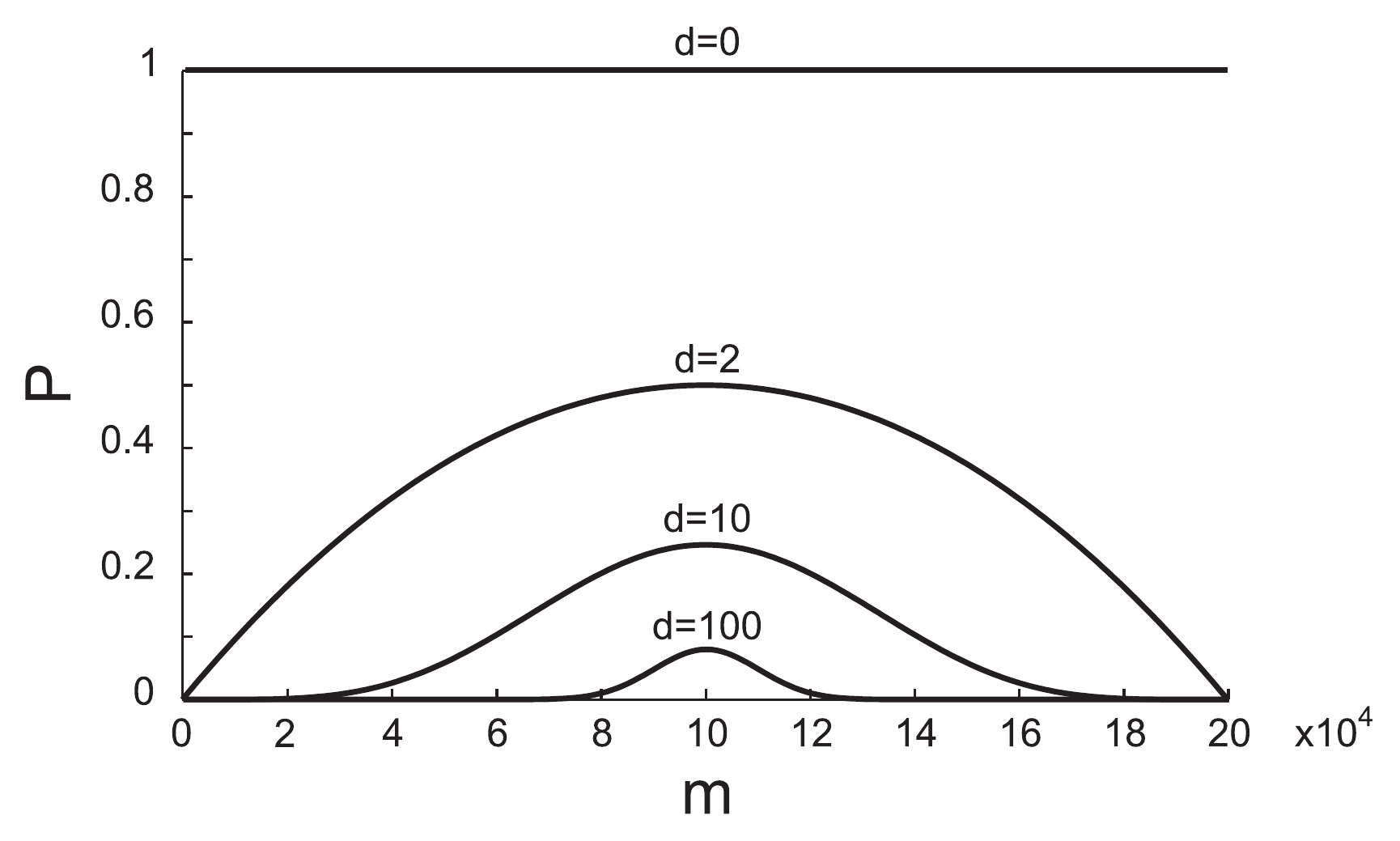} %
\caption{ Probability that binary vectors have the same Hamming weight provided certain distance between them.  $P$ is the probability that the binary vector $y$ has the Hamming weight $m$ provided $H(x,y)=d$ and $|x|=m$.  $n=20000$.}
\vspace{0.2in}
\label{fig_corol3}
\end{figure}

The result of the Corollary \ref{corol3} does not depend on the value of the threshold $\theta$. It 
characterizes properties of binary vectors {\it{per se}}. In Figure~\ref{fig_corol3}, the probability from the corollary is calculated for $n=20000$  (a typical number of inputs to a cortical neuron) depending on the number $m$ of ones (activated synapses) in the pattern and Hamming distance $d$. Note that the curves do not represent probability density functions.  For $d=0$,  $x=y$  the curve is the horizontal line $Prob=1$. The probability from Corollary \ref{corol3}  is symmetric about $m=n/2$ and has its maximum at this value. Indeed, according to an urn model for the hypergeometric distribution, the calculated probability is the probability of having an equal number of black and white balls in a sample of $d$ balls picked at random from an urn that has $m$ black and $n-m$ white balls. Accordingly, the probability is the greatest when the number of black and white balls is the same, $m=n/2$ (for even $n$) or differs by one (for odd $n$). 



\end{document}